\documentclass[]{aa}
\usepackage{graphicx}
\begin{document}
 
\title{The evolution of the Milky Way from
its  earliest phases : constraints on stellar nucleosynthesis}
\author{Patrick Fran\c cois \inst{1,2},
Francesca Matteucci\inst{3} ,
Roger Cayrel \inst{1},
Monique Spite \inst{4},
Fran\c cois Spite \inst{4} and
Cristina Chiappini \inst {5}}
\authorrunning{Fran\c cois et al.}
\titlerunning{ The evolution of the Milky Way}
\institute{
Observatoire de Paris/Meudon GEPI 61 Avenue de l'Observatoire Paris
\and 
Visiting scientist at Euopean Southern Observatory, Karl Schwarzschild Strasse 2,
D-85748, Garching, Germany
\and 
Dipartimento di Astronomia, Universit\'a di Trieste, via G.B. 
Tiepolo 11, I-34131 Trieste, Italy 
\and 
Observatoire de Paris-Meudon, GEPI,
F-92195 Meudon Cedex, France 
\and 
I.N.A.F. Osservatorio Astronomico di Trieste  via G.B. 
Tiepolo 11, I-34131 Trieste, Italy 
           }

\date{Received 1 August 2003 / Accepted 1 March 2004}

\abstract{We computed the evolution of the abundances of 
O, Mg, Si, Ca, K, Ti, Sc, Ni, Mn, Co, Fe and Zn in the Milky Way. We made use of the most widely adopted nucleosynthesis 
calculations  and compared the model results with observational data 
with the aim of imposing constraints upon stellar yields. 
To best  fit  the data in the solar neighborhood, when adopting the Woosley and Weaver (1995) yields for massive
 stars and the Iwamoto et al. (1999) ones 
for type Ia SNe, it is required that: i) the Mg yields should be increased in stars with masses from 11 to 20 $M_{\odot}$ and
 decreased in masses larger than 20$M_{\odot}$. The Mg yield should be also increased in SNe Ia. ii) The Si yields should be 
slightly increased in stars above 40$M_{\odot}$, whereas those of Ti should be increased between 11 and 20 $M_{\odot}$
 and above
 30$M_{\odot}$.  iii) The Cr and Mn yields should be increased in stars with masses in the range 11-20 $M_{\odot}$, iv) 
the Co 
yields in SNe Ia should be larger and smaller in stars in the range 11-20$M_{\odot}$, v) the Ni yield
 from type Ia SNe should be
 decreased, vi) the Zn yield from type Ia SNe should be increased.
 vii) The yields of O (metallicity dependent SN models), Ca, Fe,
 Ni, and Zn (the solar abundance case) in massive stars from
Woosley and Weaver (1995) are the best to fit the abundance patterns of these elements since they do not need any changes. 
We also adopted  the yields by Nomoto et al. (1997) and Limongi and Chieffi (2003) for massive stars and discuss the
 corrections required in these yields in order to fit the observations. Finally, the small spread in the [el/Fe] ratios 
in the metallicity range from [Fe/H]=$-$4.0 up to $-$3.0 dex (Cayrel et al. 2003)
is a clear sign that the halo of the Milky Way was well mixed even in 
the earliest phases of its evolution.
}

\maketitle

\keywords{stars:abundances -  galaxy:evolution}

\section{Introduction}
Abundances and abundance ratios in very metal poor stars ([Fe/H] $<$ $-$3.0 dex) are fundamental tools to understand the 
earliest phases of the evolution of the Milky Way as well as high redshift objects.
Moreover, from the study of the [el/Fe] versus [Fe/H] relations one can infer very important constraints on stellar nucleosynthesis
 calculations,
 stellar lifetimes and the star formation history in galaxies (see Matteucci, 2001). In particular, the study of very metal poor stars
 in the Galaxy allows us to understand the nucleosynthesis in massive stars
and thus impose constraints on 
stellar models. Very few calculations are available for stars in the range 30-100 $M_{\odot}$ and one has to extrapolate 
 for the yields in this mass range when calculating the chemical evolution of galaxies. For this reason it is
 of paramount importance to derive accurate abundances and abundance ratios in the extremely metal poor stars and this has become
 possible only recently 
by means of high resolution spectrographs such as UVES installed  at the VLT.
Cayrel et al. (2003) have derived the abundances of several $\alpha$ and 
Fe-peak elements for a sample of very metal poor giants ( [Fe/H] from $-$4.0 to 
$-$3.0 dex) thus allowing us
to test chemical evolution models in this metallicity range never 
reached before.
Several models of chemical evolution of the Milky Way are now available in 
the literature (Chiappini et al., 1997; Portinari and Chiosi 1999; 
Goswami and Prantzos 2000; 
Chang et al. 1999; Alib\`es et al. 2001) and they all share some 
important features such as the relaxation of the instantaneous recycling 
approximation (I.R.A.), which is fundamental to follow the evolution 
of elements produced on long timescales, and the inclusion of type Ia SNe 
following the original prescriptions of Greggio and Renzini (1983) and 
Matteucci and Greggio (1986). These models can reproduce the majority of the properties of the solar
 neighborhood and the whole disk, in particular they 
all agree that the disk of the Milky Way should have formed inside-out and
that the local disk assembled on a timescale of the order of 6-7 Gyr.
These models also reproduce the abundance gradients observed along the 
Galactic disk as well as the distribution of gas and star formation.
In this paper we plan to adopt an updated version of the two-infall model of Chiappini et al. (1997)
and include several different stellar yields published so far to reproduce the most recent data
 on metal poor stars as well as the data for all metallicities. We will adopt a selected and accurate data sample.
>From the comparison between theory and observations we derive 
strong constraints upon the stellar yields which
are still affected by large uncertainties. Especially uncertain are the yields of Fe-peak elements owing to the 
uncertain mass-cut, explosion energies, neutron fluxes mixing  and possible fall-back of the expelled material.
 The comparison between the new data at very low metallicity and the detailed chemical evolution models will allow us
 to suggest how to correct the yields and gain information on the nucleosynthesis of high mass stars. A similar 
approach was presented in Argast et al. (2002) where the authors tried to constrain the yields of O, Mg and Fe by fitting
 observations. However, they focused on a possible scatter present in the abundance ratios for [Fe/H] $< $-$2.5$.
 The new data for very metal poor stars 
imply only a small scatter, thus suggesting that the halo was indeed well mixed and allowing  one to impose
 constraints on the nucleosynthesis of massive stars up to 30-35 $M_{\odot}$  and  thus extend the work of 
Argast et al. (2002),
who could impose constraints only on the range 10-20$M_{\odot}$ owing to the lack of data at very low metallicities. However
,  by means of chemical evolution models one is able to impose constraints on the whole range of
 stellar masses up to $100 M_{\odot}$ since the most massive stars contribute to the solar chemical abundances even
 if we cannot impose precise constraints on their production ratios.
To impose such constraints we would need to have measured abundance ratios for [Fe/H] $<-$ 4.0 dex (Audouze \& Silk 1995, Ryan 1996).
The paper is organized as follows: in section 2 we describe the data sample,
in section 3 the adopted chemical evolution model, in section 4 the nucleosynthesis prescriptions. In section 5 we
show the comparison between models and observations and, in section 6, some conclusions are drawn.

\section{Observational data}
We have adopted  a data sample for stars in the solar neighborhood spanning a metallicity range from $-$4 to solar.
In particular, for the very metal poor stars ([Fe/H] between $-$4 and -3), we have considered the very recent results from
the UVES Large Program `` First Stars'' (Cayrel et al. 2003). This sample is made up of 35 extremely metal-poor
 giants selected  in the HK survey (Beers et al. 1992 and Beers et al.1999). The analysis has been made in a systematic and 
homogeneous way, from very high quality data, so that reliable trends of the abundances may be derived.
 It is important to note that the Beers et al. data (1992) is based on  a very large survey (1940 $deg^{2}$). 
The detection of very metal 
poor stars is based on the strength  of the H and K lines.  
The halo stars are rather uniformely distributed over the sky.
The selection of the sample of Cayrel's stars is based on the estimation
 of their metallicity and their  magnitude.
 The kinematics of these stars is such that 
they were born at very different places in the galactic halo.
So, there is no possibility for  a selection bias as could be found for 
a sample of bright disk dwarf stars located in the solar vicinity
for which similar birthplaces and kinematics may be found. 

Norris et al. (2001) introduced a factor F (defined as R x (S/N) / $\lambda$, where R is the resolution of the spectrum and
S/N the signal to noise ratio), representative of the quality of the observational data. They estimated that
F values larger than 500 are now required to make progress in the understanding of the chemical history of our Galaxy.
The data used in this study have F values ranging between 850 and 3250 , which are larger than the values from other studies.
These observations provide us with abundance ratios of unprecedented accuracy in this metallicity range.

For the abundances in the remaining range of [Fe/H], we adopted already published data in the literature from various sources :
Stephens (1999), Carney et al. (1997), Nissen and Schuster (1997), Fulbright (2000), Gilroy et al. (1998), 
Gratton \&  Sneden (1988; 1994), Ryan et al. (1991), Carretta et al. (2002), Edvardsson et al. (1993), McWilliam et al. (1995), 
Nissen et al. (2002), Matteucci et al. (1993 and references therein). All of these data are relative to
 the solar abundances of Grevesse and Sauval (1998) with the exception of oxygen for which we adopted the 
new value of  Allende Prieto
et al. (2002).

\section {The chemical evolution model for the Milky Way}

The model for the Galaxy assumes two main infall episodes for the 
formation of the halo-thick disk, and the thin-disk, respectively. 
The timescale for the formation of the thin disk is much longer than that of the halo, 
implying that the infalling gas forming the thin disk comes not only from the halo but rather mainly 
from the intergalactic medium (e.g. Chiappini et al. 1997). The timescale for the formation of the thin disk is assumed to 
be a function of the galactocentric distance, leading to an inside-out picture for the Galaxy 
disk build-up, according to the original suggestion of Matteucci and Fran\c cois
(1989). The two-infall model differs from other models in the 
literature  in two 
aspects: it considers an almost independent evolution between the halo and thin disk components 
(see also Pagel and Tautvaisiene  1995), and it assumes a threshold in the star 
formation process (see Kennicutt 1989, 1998; 
Martin \& Kennicutt 2001). The model well reproduces the majority of 
observational constraints 
about the abundances of heavy elements both locally and in the whole disk.

If $G_i$ is the mass fraction of gas in the form of an element $i$, 
we can write the main equations of the model as:

\begin{eqnarray}
 & & \dot G_i(t)  =  -\psi(r, t)X_i(r, t)\nonumber \\
& & + \int_{M_{L}}^{M_{Bm}}\psi(t-\tau_m)
Q_{mi}(t-\tau_m)\phi(m)dm\nonumber \\ 
& & + A\int_{M_{Bm}}^{M_{BM}}
\phi(M_{B}) \cdot  \\
& & [\int_{\mu_{min}}^{0.5}f(\mu)\psi(t-\tau_{m2}) 
Q_{mi}(t-\tau_{m2})d\mu]dM_{B}\nonumber \\ 
& & + (1-A)\int_{M_{Bm}}^
{M_{BM}}\psi(t-\tau_{m})Q_{mi}(t-\tau_m)\phi(m)dm\nonumber \\
& & + \int_{M_{BM}}^{M_U}\psi(t-\tau_m)Q_{mi}(t-\tau_m) 
\phi(m)dm\nonumber   + X_{A_{i}} A(r,t) 
\end{eqnarray}

The star formation rate (SFR) adopted 
here is: 
\small
\begin{eqnarray}
\psi(r, t) = \nu(t)\Big(\frac{\Sigma(r,t)}{\Sigma(r_{\odot},t)} \Big)
^{2(k - 1)}\Big(\frac{\Sigma(r,t_{Gal})}{\Sigma(r,t)} \Big)^{k-1}
G_{gas}^{k}(r,t)
\end{eqnarray}
\normalsize
where $\nu(t)$ is the efficiency of the star formation process.
$\Sigma(r,t)$ is the total  
surface mass density at a given radius $r$ and given time $t$, 
$\Sigma(r_{\odot}, t)$ is the total 
surface mass density at the solar position, $G_{gas}(r, t)$ is the 
surface gas density normalized 
to the present time total surface mass density in the disk 
$\Sigma_{D}(r,t_{Gal})$, 
t$_{Gal}$ = 13 Gyr is the age of the Galaxy, r$_{\odot}$ = 8 kpc is 
the assumed solar galactocentric 
distance 
(see Reid 1993).

The gas surface density exponent, $k$, 
is set equal to 1.5, to 
ensure a good fit to the observational constraints in the solar vicinity. 
This value is also 
in agreement with the observational results of Kennicutt (1998), and 
with N-body simulation 
results by Gerritsen \& Icke (1997). The star formation efficiency is set to 
$\nu$ = 2 Gyr$^{-1}$, for the Galactic halo, whereas it is $\nu$ = 1 
Gyr$^{-1}$ for the disk; 
this is to ensure the best fit to the observational features in the solar 
vicinity. The star formation
rate becomes zero when the gas surface density drops below a certain 
critical threshold (see Chiappini et al. 2001 for details).  The assumption of such a threshold density
naturally produces  a hiatus  in the SFR
between the halo-thick disk phase and the thin disk phase. This 
discontinuity in the SFR seems to be observed in the 
[Fe/O] vs. [O/H] 
(Gratton et al. 2000) and in the [Fe/Mg] vs. [Mg/H] (Fuhrmann 1998) plots.
The initial mass function (IMF) is that of Scalo (1986) and is 
assumed to be constant in 
time and space.

The SNeIa rate has been computed following 
Greggio and Renzini (1983) and 
Matteucci and Greggio (1986) and is expressed as:
\begin{eqnarray}
R_{SNeIa}=A\int_{M_{B_{m}}}^{M_{B_{M}}}\phi(M_{B})\int_{\mu_{m}}^{0.5}f(\mu)\psi(t - \tau_{M_{2}})d\mu~dM_{B},
\end{eqnarray}
where $M_{2}$ is the mass of the secondary, $M_{B}$ is the total mass 
of the binary system, 
$\mu = M_{2}/M_{B}$, $\mu_{m}=max\big\{M_{2}(t)/M_{B}, 
(M_{B}-0.5M_{B_{M}})/M_{B}\big\}$, 
$M_{B_{m}}$ = 3 $M_{\odot}$, $M_{B_{M}}$ = 16 $M_{\odot}$. 
The IMF is represented by
$\phi(M_{B})$  and
refers to  the total mass of the binary system for the computation of the 
SNIa rate, $f(\mu)$ is the 
distribution function 
for the mass fraction of the secondary, 
$f(\mu)=2^{1+\gamma}(1+\gamma)\mu^{\gamma}$, with 
$\gamma$ = 2;  $A$ = 0.05 is the fraction of systems with total 
mass in the appropriate range, 
which give rise to SNIa events. This quantity is fixed by reproducing the 
observed SNe Ia rate at the present epoch (Cappellaro et al. 1999; see also Madau et al. 1998). 

The term $A(r,t)$ represents the accretion term and is defined as:
\begin{equation}
A(r,t)= a(r) e^{-t/ \tau_{H}(r)}+ b(r) e^{-(t-t_{max})/ \tau_{D}(r)}
\end{equation}
$X_{A_{i}}$ are the abundances in the infalling material, which is assumed to be primordial,  while $t_{max}=1$Gyr
is the time for the maximum infall on the thin disk, $\tau_{H}= 2.0 Gyrs$
is the time scale for the formation of the halo thick-disk and $\tau(r)$
is the timescale for the formation of the thin disk and is a function of the galactocentric distance (formation inside-out, 
Matteucci and Fran\c cois, 1989;Chiappini et al. 2001).
In particular, we assume that:
\begin{equation}
\tau_{D}=1.033 r (kpc) - 1.267 \,\, Gyr
\end{equation}
Finally, the coefficients $a(r)$ and $b(r)$ are obtained  by imposing a fit to 
the current total surface mass density as a function of galactocentric 
distance.  In particular, $b(r)$ is assumed to be different from zero 
only 
for $t \ge t_{max}$.

\begin{table*}[!ht]
\begin{flushleft}
\caption{ The stellar yields (expressed in solar masses) as derived in this paper, namely those that produce the best 
agreement with observations. Yields for O are identical to metal dependent yields of WW95 }
\begin{tabular}{cccccc}
&&&&&\\
m($M_{\odot}$)& Mg & Si & Ca& Fe &Zn \\
\hline
&&&&&\\
11.00 &  .6440E-01& .2170E-01& .1400E-02& .8000E-01& .3510E-04 \\
12.00 &  .5740E-01& .9090E-01& .1500E-01& .5500E-01& .3360E-04 \\
13.00 &  .1148E+00& .5850E-01& .3600E-02& .1460E+00& .4500E-04 \\
15.00 &  .1869E+00& .1100E+00& .1110E-01& .1297E+00& .7280E-04\\
18.00 &  .3864E+00& .1370E+00& .6900E-02& .8300E-01& .1060E-03\\
19.00 &  .1757E+00& .2770E+00& .1380E-01& .1170E+00& .1560E-03\\
20.00 &  .2191E+00& .2880E+00& .1480E-01& .1060E+00& .3160E-03\\
22.00 &  .2912E+00& .3560E+00& .1770E-01& .2225E+00& .3240E-03\\
25.00 &  .7000E-01& .3150E+00& .1690E-01& .1500E+00& .4070E-03\\
30.00 &  .7000E-01& .3160E+00& .1340E-01& .2500E-01& .3818E-03\\
35.00 &  .7000E-01& .1120E+00& .1600E-02& .2770E-01& .4176E-03\\
40.00 &  .7000E-01& .5460E-01& .1600E-02& .2830E-01& .3710E-03\\
50.00 &  .7000E-01& .5460E-01& .1600E-02& .2830E-01& .3710E-03\\
60.00 &  .7000E-01& .5460E-01& .1600E-02& .2830E-01& .3710E-03\\
70.00 &  .7000E-01& .5460E-01& .1600E-02& .2830E-01& .3710E-03\\
80.00 &  .7000E-01& .5460E-01& .1600E-02& .2830E-01& .3710E-03\\
90.00 &  .7000E-01& .5460E-01& .1600E-02& .2830E-01& .3710E-03\\
100.00&  .7000E-01& .5460E-01& .1600E-02& .2830E-01& .3710E-03\\
&&&&&\\
\hline
\end{tabular}
\label{Table1}
\end{flushleft}
\end{table*}

\begin{table*}[!ht]
\begin{flushleft}
\caption{ The stellar yields (expressed in solar masses) as derived in this paper, namely those that produce the best 
agreement with observations.}
\begin{tabular}{cccccccc}
&&&&&&&\\
m($M_{\odot}$)& K& Sc & Ti & Cr &Mn& Co& Ni \\
\hline
&&&&&&&\\
11.00 & .4040E-04 & .2564E-05& .3300E-03& .2037E-02& .1290E-03& .1272E-03& .1670E-01\\
12.00 & .1712E-02 & .2714E-05& .3960E-03& .2091E-02& .1400E-03& .2010E-04& .7000E-02\\
13.00 & .6160E-04 & .2553E-05& .3990E-03& .1929E-02& .4490E-03& .2019E-04& .1120E-01\\
15.00 & .2544E-03 & .6751E-05& .7515E-03& .1371E-02& .4710E-03& .3030E-04& .6800E-02\\
18.00 & .1176E-03 & .2415E-05& .8400E-03& .8910E-03& .4040E-03& .4290E-04& .3760E-02\\
19.00 & .6952E-03 & .6038E-05& .8670E-03& .7980E-03& .2060E-03& .7710E-04& .4680E-02\\
20.00 & .7496E-03 & .1086E-04& .8700E-03& .5400E-03& .2130E-03& .8160E-04& .4440E-02\\
22.00 & .1200E-02 & .2530E-04& .8940E-04& .4050E-03& .2300E-03& .8820E-04& .1120E-01\\
25.00 & .4032E-03 & .2622E-05& .7500E-04& .3900E-03& .2580E-03& .7500E-03& .7590E-02\\
30.00 & .2648E-03 & .1610E-05& .1500E-03& .4500E-03& .1040E-03& .8070E-03& .3350E-01\\
35.00 & .1328E-03 & .1345E-05& .4500E-03& .4500E-03& .1330E-03& .7950E-03& .2040E-02\\
40.00 & .8960E-04 & .1621E-05& .4560E-03& .9120E-04& .1340E-03& .8250E-03& .2120E-02\\
50.00 & .8960E-04 & .1621E-05& .4560E-03& .6120E-04& .1340E-03& .8250E-03& .2120E-02\\
60.00 & .8960E-04 & .1621E-05& .4560E-03& .3120E-04& .1340E-03& .8250E-03& .2120E-02\\
70.00 & .8960E-04 & .1621E-05& .4560E-03& .3120E-04& .1340E-03& .8250E-03& .2120E-02\\
80.00 & .8960E-04 & .1621E-05& .4560E-03& .3120E-04& .1340E-03& .8250E-03& .2120E-02\\
90.00 & .8960E-04 & .1621E-05& .4560E-03& .3120E-04& .1340E-03& .8250E-03& .2120E-02\\
100.00& .8960E-04 & .1621E-05& .4560E-03& .3120E-04& .1340E-03& .8250E-03& .2120E-02\\
&&&&&&&\\
\hline
\end{tabular}
\label{Table2}
\end{flushleft}
\end{table*}

\section{Nucleosynthesis prescriptions}

We divide stars into three fundamental mass ranges : i)  very low mass stars ($M< 0.8
M_{\odot}$), ii) low and intermediate mass stars ($0.8 \le M/M_{\odot} \le 8$) and iii) massive stars ($M > 8 M_{\odot}$).
Very low mass stars do not contribute to the chemical enrichment but only to lock up gas. Low and intermediate mass stars 
contribute to He, $^{12}C$,
$^{13}C$, $^{14}N$ and to some s-process elements (see Travaglio et al. 1999).
Massive stars are responsible for the formation of the bulk of $\alpha$-elements (O, Mg, Ne, Si, S, Ca, Ti)  
plus some Fe and Fe-peak elements whose yields are  rather uncertain. These stars end their lives as type II supernovae.
Type Ia SNe (C-O white dwarfs in binary systems) are instead considered to be responsible for the production
of the bulk of Fe and Fe-peak elements.

In this paper we adopt the nucleosynthesis prescriptions of van den Hoek and Groenewegen (1997) for
single low and intermediate mass stars,  of Iwamoto et al. (1999) for the yields from type Ia SNe (model W7) and  the yields
of Woosley and Weaver (1995) (hereafter WW95, their case A for stars below 
30$M_{\odot}$ and their case B for stars between 30 and 40 $M_{\odot}$), 
Nomoto et al. (1997) (hereafter N97) and 
Limongi and Chieffi (2003) (hereafter LC03) for massive stars. 
While WW95 have provided yields for different initial stellar metallicities,  
those of N97 and LC03 refer only to the solar chemical composition. 
Generally, yields of primary elements, namely those elements produced 
starting  directly from the H and He through the chain of hydrostatic burnings in stars, depend only 
slightly on
the initial stellar metallicity. Therefore, we have chosen to  consider  the WW95 yields 
of primary elements that  refer to the solar chemical composition with the 
exception of oxygen (see next section).
For Zn, which has a more complex nucleosynthetic origin, being partly produced in explosive nucleosynthesis 
and partly being manufactured as an s-process element in massive stars, we adopted WW95 values for the explosive nucleosynthesis
 in massive stars  and the prescriptions of Matteucci et al. (1993) for the other components (nucleosynthesis in type Ia SNe and quiescent 
He-burning in massive stars during which weak s-processing takes place).
In particular, these authors suggested that the yield of Zn from type Ia SNe should be higher than predicted and that
 it represents the major component in the Zn production.
This element is quite important since is the best tracer of metallicity in high redshift objects such as Damped
 Lyman-$\alpha$ systems (DLA) and Lyman-break galaxies,
owing to the fact that it is only very slightly  depleted into dust.
Stellar yields, especially those of Fe-peak elements, 
are still quite uncertain since they strongly depend upon the choice of the
mass cut between ejecta and the proto-neutron stars, the explosion 
energies and 
the neutron flux mixing.
On the other hand,  the yields of elements produced during hydrostatic burnings such as O and Mg should 
be better known. However, the Mg yield is more sensitive than O to the treatment of convection used by the different authors.
In this paper, we have adopted a  stellar mass range of 0.1-100$M_{\odot}$ and since the available nucleosynthesis 
prescriptions go only until 70$M_{\odot}$ (N97) we have kept the yields constant and equal to the value corresponding 
to the largest  computed stellar mass for which the yields are available.

\section{Model results}
In Figures 1 and 2 and 3 we show the model predictions for several elements (O, Mg, Si, Ca, K, Ti,  Ni, Sc, Cr, Mn, Co and Zn),
 in particular for the relations
[el/Fe] versus [Fe/H] compared with the observational data described before.
The yields that we have adopted are those of the SNII of WW95 with initial  solar chemical composition, with the exception of oxygen.
\begin{figure}
\resizebox{\hsize}{!}{\includegraphics{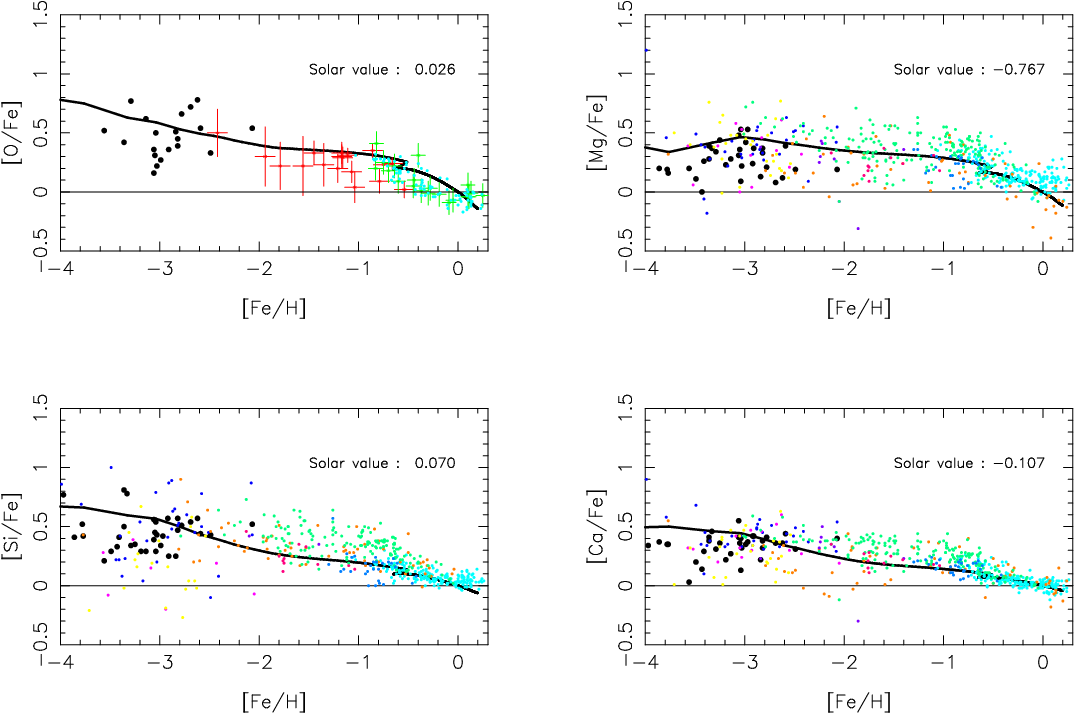}}
\caption{[el/Fe] versus [Fe/H] for several $\alpha$- elements
compared to a large compilation of data (see text). The black dots represent the new data of Cayrel et al. (2003).
 Here we have adopted the yields of WW95 for massive stars and N97
 for type Ia SNe taken as per theifa models.The model predictions are normalized to the predicted solar abundances, namely those
 predicted at 4.5 Gyr ago. In a corner of each panel we show our predicted solar abundance ratios.  
In some cases we fit the behavior of [el/Fe] vs. [Fe/H] but not the solar value, as it is the case for Mg.}
\label{Fig1}
\end{figure} 

\begin{figure}
\resizebox{\hsize}{!}{\includegraphics{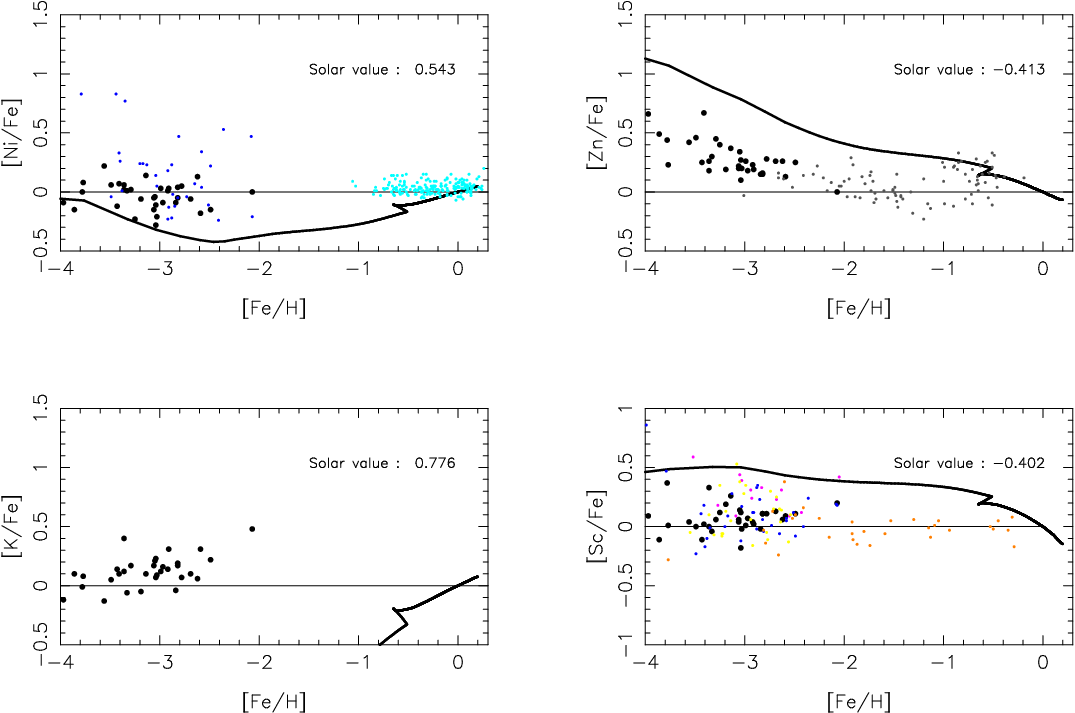}}
\caption{The same as in figure 1 for several Fe- peak elements. 
The yields are taken at as per their models.}
\label{Fig2}
\end{figure} 

\begin{figure}
\resizebox{\hsize}{!}{\includegraphics{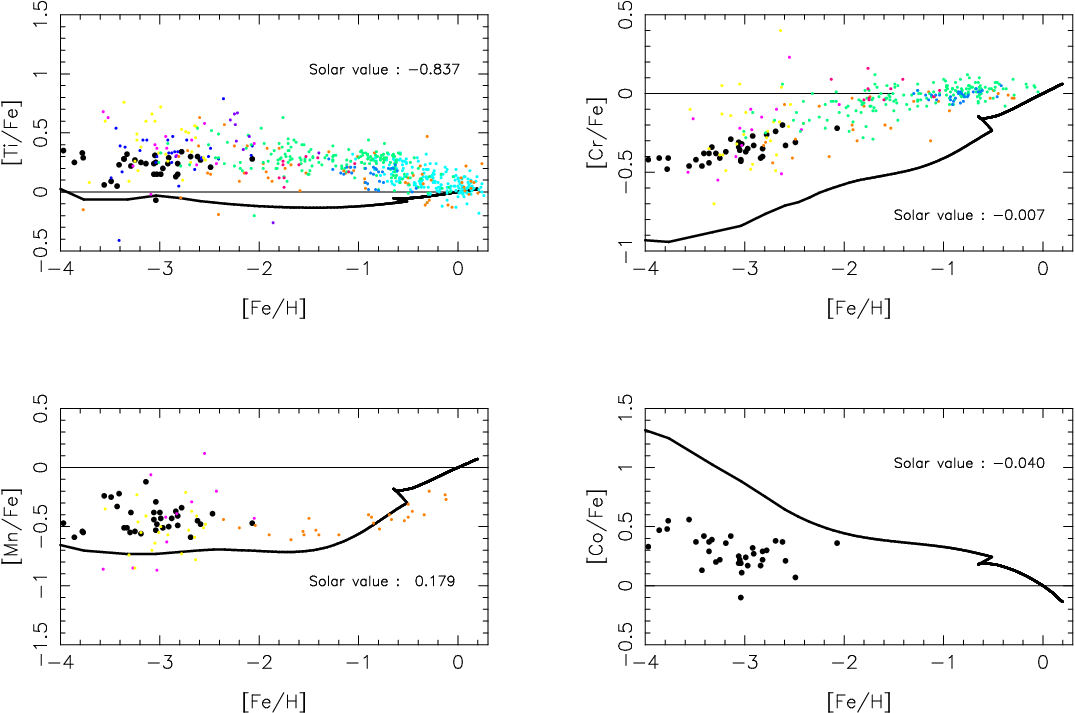}}
\caption{The same as in figure 1 for several Fe- peak elements. 
The yields are taken as per their models.}
\label{Fig3}
\end{figure}

For this element we adopted the metallicity-dependent yields of WW95  since they best fit the new data of oxygen at very low
metallicity.
Goswami and Prantzos (2000) had already published the predictions for [O/Fe] vs. [Fe/H] obtained  by using both
the oxygen yields of WW95 as a function of metallicity as well as those for 
the solar chemical composition, and although the differences between these 
two cases are generally small, especially for primary elements, it was 
evident from their figures that the metallicity-dependent case was the best to reproduce the oxygen data available then. 
For type Ia SNe we took the yields of Iwamoto et al. (1999) from  their model W7. In all our models we 
have normalized the 
predicted abundances to the predicted solar abundances, which reflect the abundances in the gas 4.5 Gyr ago.
Therefore, the predicted curve includes the 0,0 point.
In each panel of Figures 1, 2 and 3 we print in the top right the solar [el/Fe] ratio 
that we predict for the time of formation of the solar
system and relative  to the observed solar abundances (Grevesse \& Sauval, 1998 for all the elements except oxygen for 
which we adopt the newer
estimates of Holweger, 2001 and Allende-Prieto et al. 2001).
A good model of chemical evolution of the Milky Way should be able to reproduce both 
the abundance patterns and the absolute solar abundances.
As one can see from  Figures 1, 2 and 3  the O, Mg, Si and Ca behaviours
are very well fitted as are their solar values with the exception of 
the [Mg/Fe] ratio,
which is largely underestimated, owing to the too low Mg yields predicted for massive stars, a problem common to all chemical 
evolution models (e.g. Chiappini et al. 1999; Thomas et al. 1999). On the other hand, for the other elements 
(Ni, Zn, K, Sc, Ti, Cr, Mn and Co) the trends are not well reproduced and, with the exception of Co, Mn and Cr, not even the 
predicted solar abundance ratios are confirmed. Therefore, there is a clear indication that it is necessary to modify the yields, especially 
those of the Fe-peak elements. 
In Figures 4, 5 and 6  we show the predictions obtained with the yields modified ``ad hoc'' to fit the data, 
according to the prescriptions given in Figure 7,
where we show the ratios between the suggested and the published yields of WW95 for SNII as well as those of Iwamoto et al. (1999),
 model W7, for the type Ia SNe.The suggested yields  are shown in Tables 1 and 2. 

 As an example we examine the treatment adopted for the yields of K which was applied  to fit  the variation of [K/Fe] vs [Fe/H]
found in the halo stars. Figures 2, 5  and 7 for K shows what treatment has been done
to fit the data points. Using the yields for SNII (WW95) and SNIa (Nomoto) gives a 
too high value of [K/Fe] at the solar birth ([K/Fe]=0.776). In Figure 2, an increasing slope
from metal poor stars to solar metallicity reveals that the production ratio SNII/SNIa is much too 
high (we have a rather similar case for Cr,Mn and Ni). 
Therefore, a decrease of this ratio by a factor of 8  allows us  to get a decreasing [K/Fe] as a function of increasing
metallicity. The final adjustment for the SNII yields is used to get a good [Fe/H] value at solar birth.
As [K/Fe] is constant in the metal-poor stars, there is no need to change the SNII yields as a function of mass.

We have performed a series of tests to evaluate the sensitivity of the yields found in Table 1 and Table 2  to different 
assumptions of  the models of chemical evolution of the Galaxy. First, we changed the efficiency of the Star Formation
 Rate by a factor of 2. 
It changes the absolute abundances and then affects the Fe abundance found at solar birth. 
The impact on the abundance ratios is less than 0.03 dex.  
We also estimated the impact of the change of the IMF Scalo coefficient by $\pm 0.2 $ 
and found that the impact on the abundance ratio is less than 0.10 dex. We  performed a test on the relative sensitivity of the yields
of the elements for different mass ranges. We divided the masses in 3 mass ranges. For each element and each mass range we multiplied or
 divided
the yields of Tables 1 and 2 iteratively and then determined a factor such that the yields of an element in a given mass range can be 
multiplied
or divided by this factor without giving abundance ratios that are not in agreement with the observational data. 
These factors are given  in Table 3. Fe is not included in this table as it has been used as a reference element in computing  the variations
of the other yields.

\begin{table*}[!ht]
\begin{flushleft}
\caption{Sensitivity factors for the yields found in SNII :
 these factors give the amount by which the yields given in Table 1 and Table 2 can be multiplied or divided
such that the results of the model still give  fair fits to the data. }
\begin{tabular}{cccc}
\hline
Elt &    11 to 19  M$_{\odot}$   &    20 to 30 M$_{\odot}$ &    35 to 100 M$_{\odot}$ \\
\hline
O     &     2    &        2     &       2   \\
Mg    &     2    &        2     &       2   \\
Si    &     3    &        3     &       3   \\
Ca    &     2    &        2     &       5   \\
Zn    &     5    &        2     &       2 \\ 
K     &     2    &        2     &       2 \\
Sc     &    2    &        2     &       2 \\
Ti     &    2    &        2     &       2 \\
Cr     &    2    &        2     &       5 \\
Mn     &    2    &        2     &       2  \\
Co     &    2    &        5     &       2  \\
Ni     &    2    &        2     &       2  \\       
\hline
\end{tabular}
\label{Table3}
\end{flushleft}
\end{table*}

As it is evident from figure 7, the yields that did not need any revision relative to the prescriptions of WW95 are those 
of O (computed as a function of metallicity),
Fe, Ca, Zn, Ni and  K  (corresponding to the solar chemical composition).
For all other elements ( Mg, Si, K, Ti, Sc, Cr, Mn, Co) some variations of the WW95 yields are required.
In particular, the
Mg yields predicted by the available nucleosynthesis calculations for massive stars need revision to reproduce 
the solar abundance of this element. 
Starting from the calculations of WW95,
one needs to assume that the Mg yields from stars in the range 11-20 $M_{\odot}$ should be roughly a factor
 of 7 higher than predicted whereas those from stars larger than 20$M_{\odot}$ should be lower than predicted (a factor of 2 
on average). At the same time, to preserve the observed pattern of [Mg/Fe] vs. [Fe/H]
one needs also to increase the  Mg yields from type Ia SNe by a factor of 5. 
The yields of K should be multiplied by a factor of 0.8 over the whole mass range
in the case of WW95.
As we can see in Figures 8 and 9, where
we show how one needs to modify the yields to reproduce the observations if one uses the yields of N97 and LC03, respectively,
 the same type of corrections should be applied to the other examined type II SN yields.
We recall that these sets of yields (N97 and LC03) were computed only for  solar chemical composition.
Also in these cases, the Mg yields need to be increased below 20$M_{\odot}$, 
and decreased above  20$M_{\odot}$, although for the yields of LC03 the required increase is smaller than in WW95.
Therefore, one can conclude that the predicted Mg yields, 
either from type Ia or type II SNe below $20M_{\odot}$,  are too low. 
Another $\alpha$-element which seems largely underestimated over the whole mass range, in the three sets of yields (WW95, N97, LC03),
 is titanium, especially in the range 11-20$M_{\odot}$.
Concerning Si, only the yields of the very massive stars ($M> 40 M_{\odot}$) should be increased by a factor of 2, in the case of
 WW95 yields, whereas in the case of N97 they should be increased, especially in the range 15-25$M_{\odot}$ and lowered for more
 massive stars.
 The yields of Si  LC03 should be only slightly increased in the range 20-30$M_{\odot}$  and decreased by a factor
of $\sim 4$ for larger masses. 

It is not easy to envisage how to obtain the requested yields from nucleosynthesis since the $\alpha$-elements O, Mg, Si and Ca are
 produced in different 
nuclear environments: O in He-burning, Mg in C-burning and Si and Ca in explosive O-burning and in explosive incomplete Si-burning.
In addition, the amount Si and Ca can depend on the choice of the mass cut whereas O and Mg do not. 
Certainly the rate of the $^{12}C(\alpha, \gamma)^{16}O$ reaction is a very important parameter in determining not only the amounts
 of O and Mg but also the ratio between elements such as Ne, Na, Mg and  Al relative to Si, S, Ar and Ca (Imbriani et al. 2001).
A recent paper by Rauscher et al. (2002) adopting  up to date  experimental and theoretical nuclear data, new opacity tables and updated 
nuclear reaction network and considering stellar evolution with mass loss provided Mg yields higher than the previous ones of WW95 but 
still not high enough to reproduce the Mg data (they should still be increased by a factor of 3). On the other hand, their yields for
 a 25 $M_{\odot}$ supernova are too high by a factor of 2.
The yields of Cr and Mn should be
higher in the mass range 13-30$M_{\odot}$ and slightly lower for more massive stars than predicted by WW95. Finally, the yields of Co
 of WW95 should be lowered
in the range 11-22$M_{\odot}$ and increased for more massive stars, and the yield of Sc should be generally increased in the same mass
 range and
decreased for larger masses.
The yields of Fe from WW95 relative to the case of solar chemical composition are very good, whereas if one adopts the WW95 Fe yields
 as functions of metallicity they tend to overproduce Fe which then needs to be lowered by a factor of $\sim 2$.
Concerning the Fe yields from N97 and LC03, in both cases Fe needs to be slightly
increased in the range 13-25$M_{\odot}$ (LC03) and in the range 18-35$M_{\odot}$ (N97).
We recall here that Fe and Fe-peak elements (Sc, Fe, Ni, Co, Zn) are strongly dependent on the chosen mass cut, and different yields
 for these elements can be obtained under different assumptions on the mass-cut (although not independently), as shown by 
Nakamura et al. (1999).

\begin{figure}
\resizebox{\hsize}{!}{\includegraphics{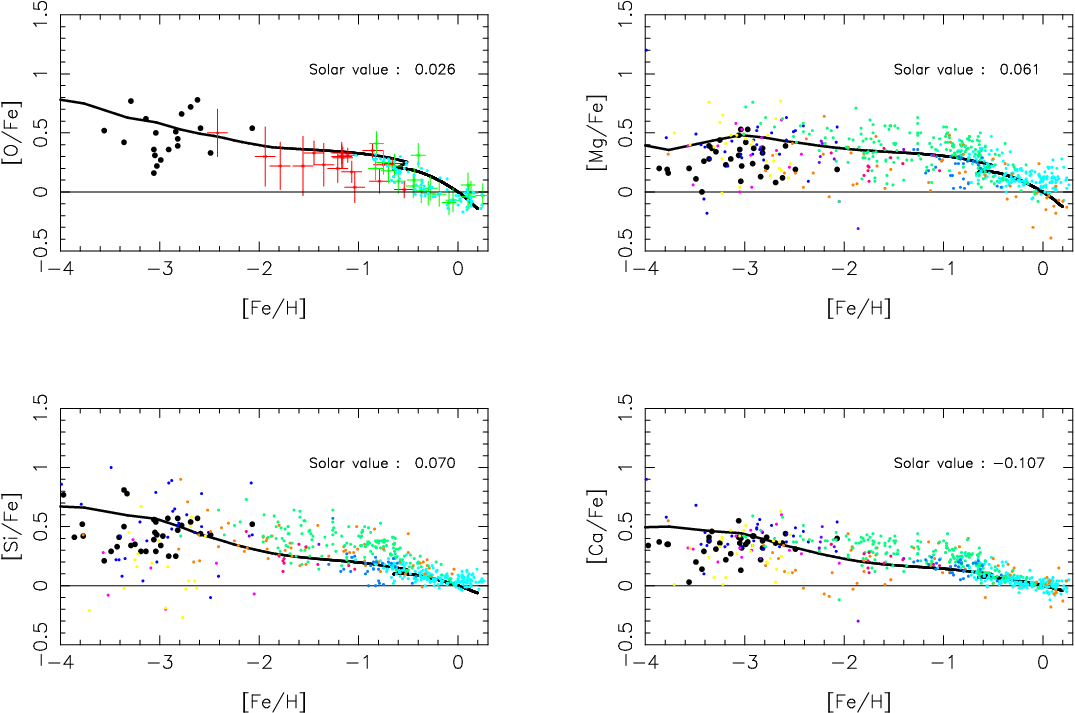}}
\caption{[el/Fe] versus [Fe/H] for several $\alpha$-elements, as predicted by adopting the corrected yields, 
compared with a large data sample (Table 1). The model predictions are normalized to the predicted solar abundances, namely 
those predicted for the gas at 4.5 Gyr ago.
 As one can see, the model reproduces now both the behavior of 
abundance ratios and the solar abundances.}
\label{Fig4}
\end{figure}

\begin{figure}
\resizebox{\hsize}{!}{\includegraphics{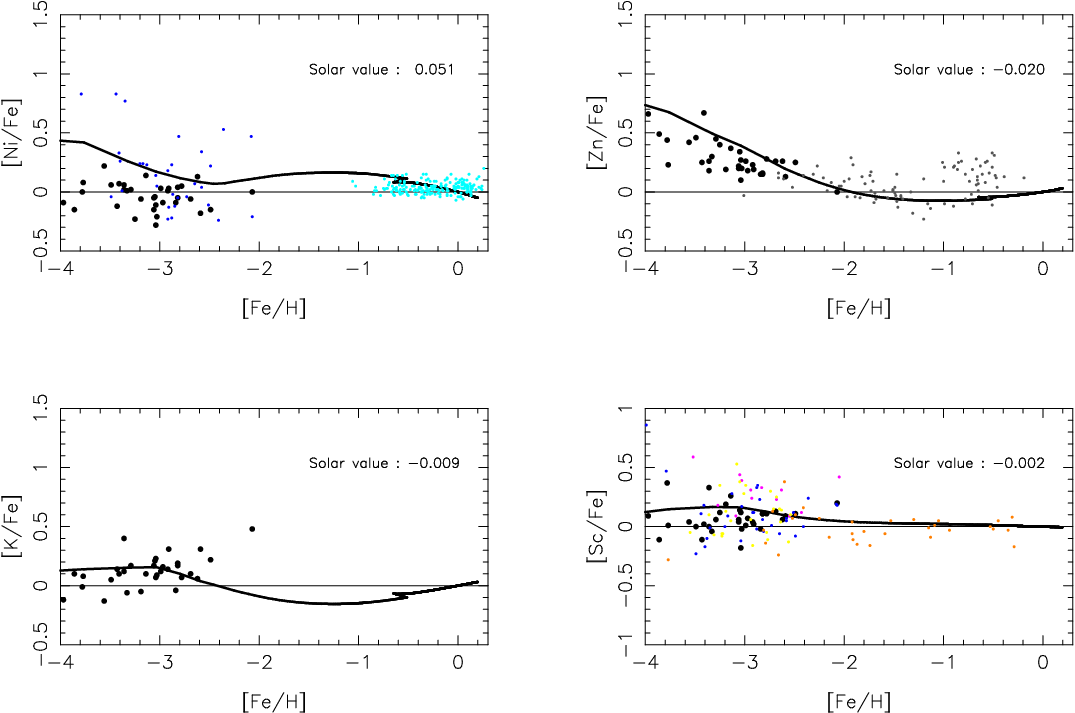}}
\caption{The same as figure 2 for some Fe- peak elements.}
\label{Fig5}
\end{figure} 

\begin{figure}
\resizebox{\hsize}{!}{\includegraphics{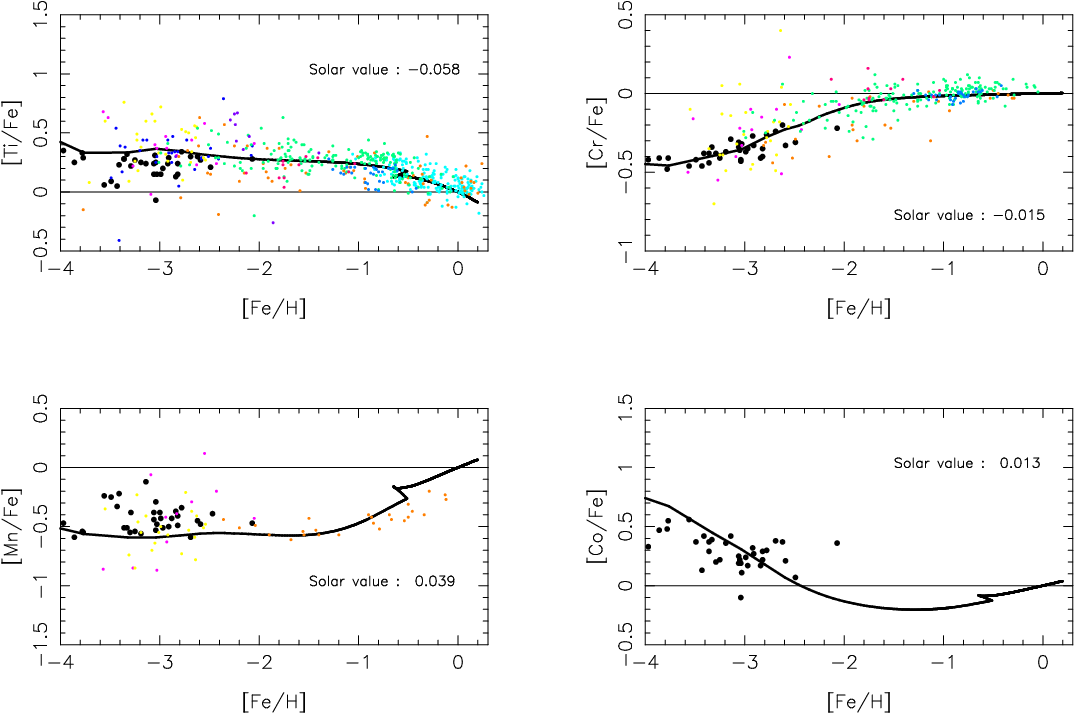}}
\caption{The same as figure 2 for some Fe- peak elements.}
\label{Fig6}
\end{figure} 

The yields of Mn should be increased only in the range 13-18 $M_{\odot}$
and decreased for stars $>30 M_{\odot}$ by a factor of $\sim 2.5$ relative to the WW95 yields.
For the massive star calculations of N97, Mn and Cr should be decreased for masses $M \ge 25 M_{\odot}$ whereas Sc should be 
increased by huge factors in the range 13-25 $M_{\odot}$.

By examining Figures 7, 8 and 9 we can conclude that
the yields of WW95 need less correction of the number of elements, whereas the yields of N97 need 
corrections for all the studied isotopes as well as those of LC03. However, the yields of LC03 generally need 
relatively small corrections with the exception of K and Ni.
Concerning the yields from type Ia SNe (Iwamoto et al. 1999, model W7), those which need a revision are  Mg (should be higher), 
Ti (higher), Sc (almost a factor of 100 higher),  Zn (higher), Co (higher), K (lower) and Ni (lower).
The nucleosynthesis of Zn has already been studied by Matteucci et al. (1993)
who concluded that the Zn yields from type Ia SNe should be higher by a factor of $\sim 10$ relative to model W7 of 
Nomoto et al. (1984).

\section{Summary and conclusions}
In this paper we compared theoretical predictions about the [el/Fe] vs. [Fe/H]
trends in the solar neighborhood for
several chemical elements (O, Mg, Si, Ca, K, Ti, Sc, Ni, Mn, Co, Fe and Zn)
with high quality spectroscopic data. In particular, we considered the very recent abundance determinations by Cayrel et al. (2003)
 in the metallicity range [Fe/H]=$-$3.0 down to $-$4.0 dex. These data allow us to impose constraints both on the element production
over the whole galactic lifetime and on all masses contributing to chemical enrichment. They allow us to suggest precise 
yield ratios for stars up to $\sim 35 M_{\odot}$, since for the more massive ones we would need observed abundance ratios for
 [Fe/H] $<-$ 4.0. 
For  metallicities between  [Fe/H] = $-$3.0 and $-$4.0 stars with masses between 
$\sim$ 30 and 35  $M_{\odot}$  contribute to the Galactic chemical enrichment.
The comparisons performed in this paper allow us also to infer constraints on the nucleosynthesis in type Ia SNe since we have
 considered the behavior of the [el/Fe] ratios over the whole [Fe/H] range.

\begin{figure}
\resizebox{\hsize}{!}{\includegraphics{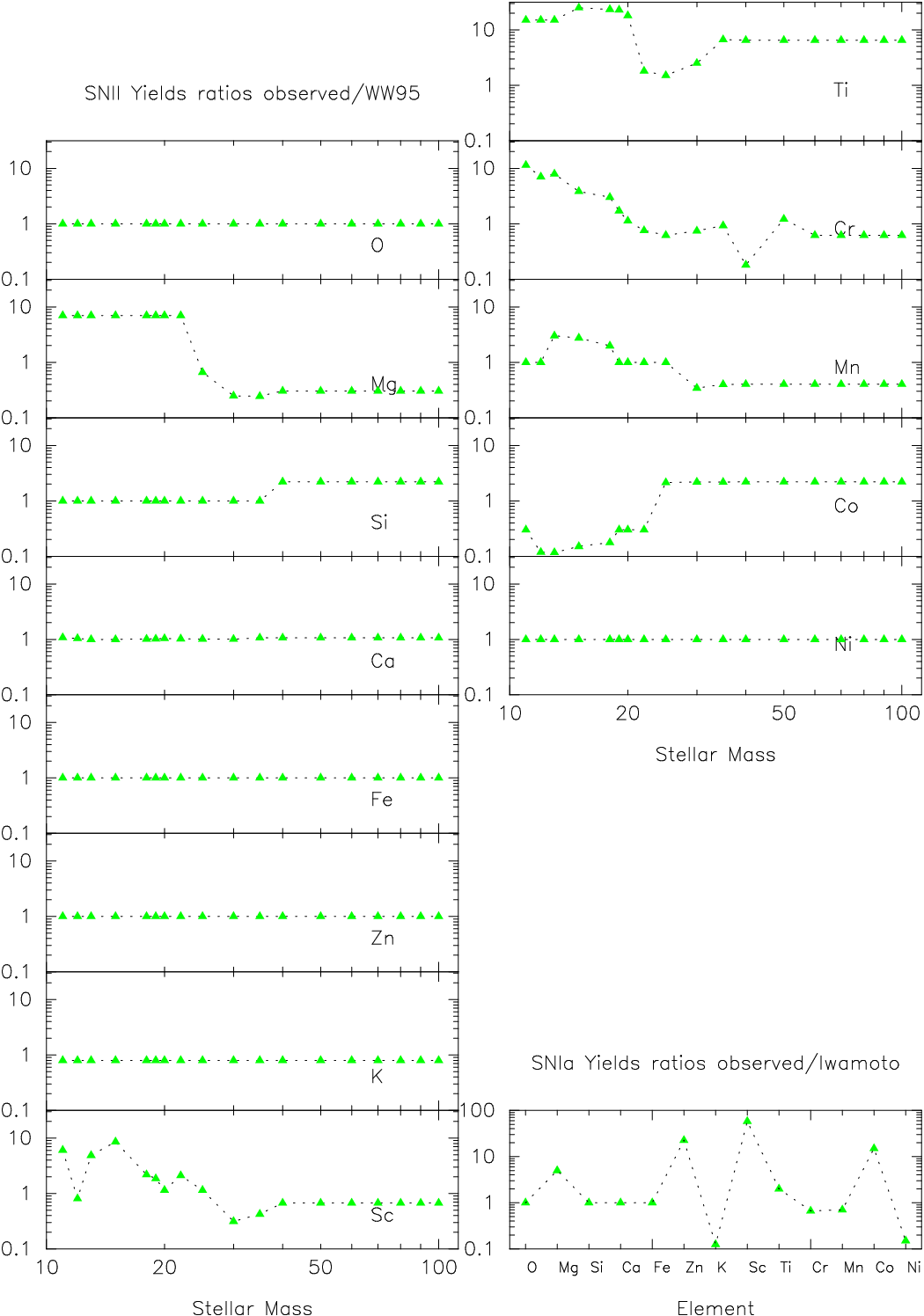}}
\caption{Ratios between the yields we adopted to obtain the best fit to the data and the yields of WW95 for massive stars and 
those of Iwamoto et al. (1999) for the nucleosynthesis in type Ia SNe. Notice that the oxygen yields are those of WW95 as a
 function of metallicity whereas all the other yields of WW95 refer to the solar chemical composition.}
\label{Fig7}
\end{figure} 

\begin{figure}
\resizebox{\hsize}{!}{\includegraphics{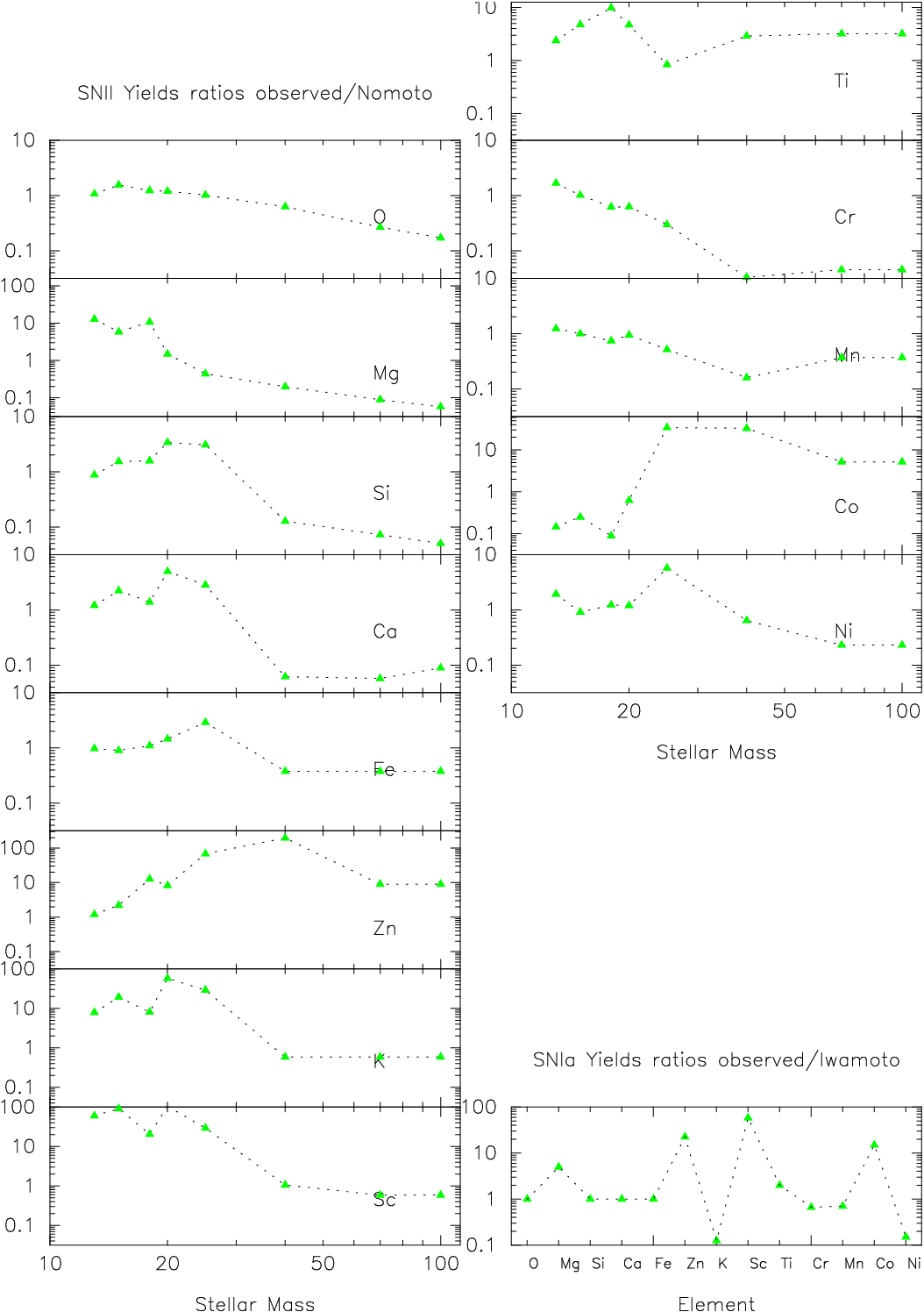}}
\caption{Ratios between the yields we adopted to obtain the best fit to the data and the yields of N97 for massive stars and
 those of Iwamoto et al. (1999)
for type Ia SNe.}
\label{Fig8}
\end{figure}

\begin{figure}
\resizebox{\hsize}{!}{\includegraphics{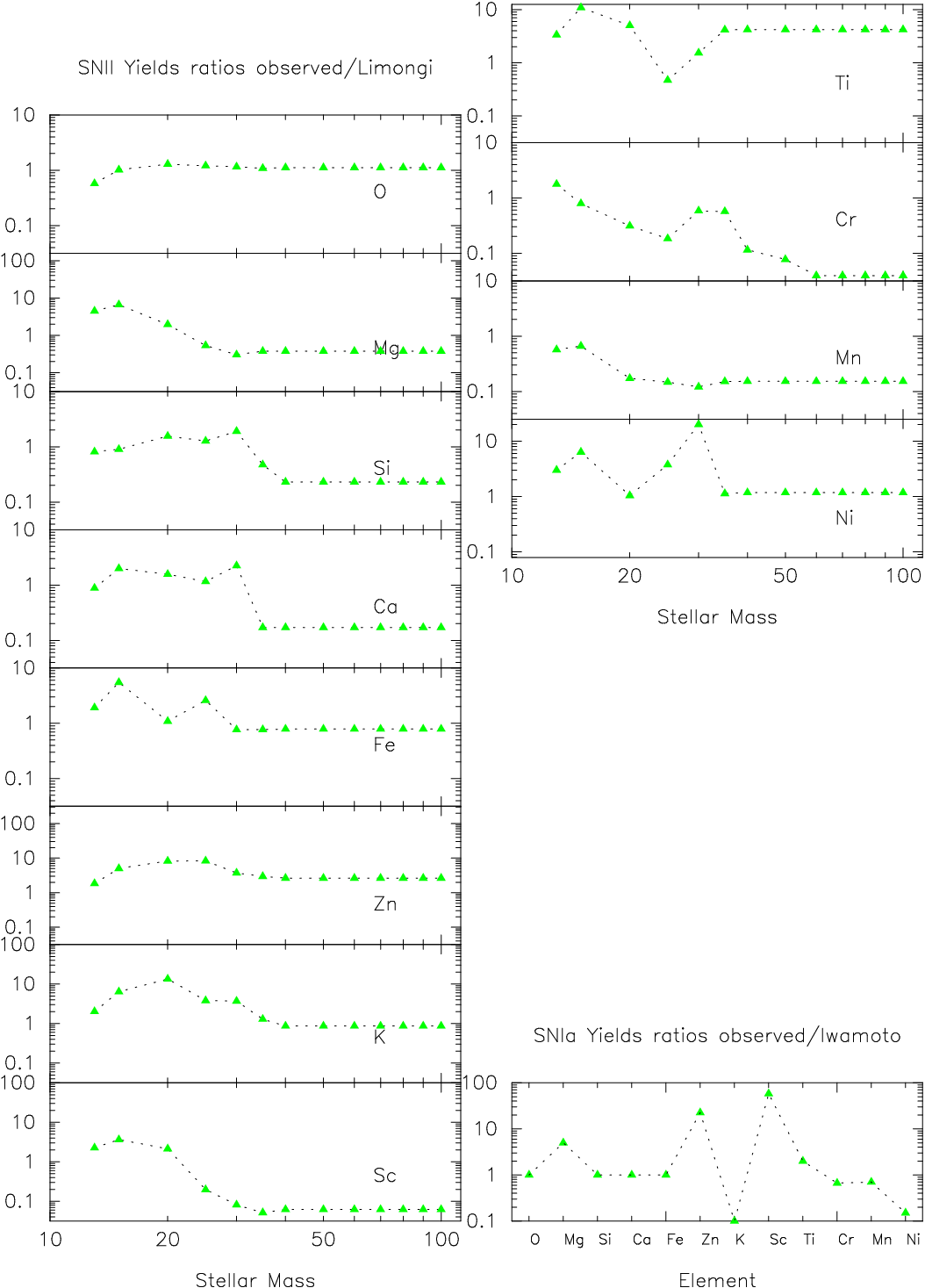}}
\caption{Ratios between the yields we adopted to obtain the best fit to the data and the yields of LC03 for massive stars and
 those of Iwamoto et al. (1999) for the nucleosynthesis in type Ia SNe.}
\label{Fig9}
\end{figure}

Our main conclusions can be summarized as follows:
\par
\begin{itemize}
\item A homogeneous model for the early halo chemical evolution is able to fit the observations, for metallicities [Fe/H] $<$ -3.0 dex
 , since the data for [el/Fe] at these low metallicities show a  small spread (typically 0.2 - 0.3 dex).

\item The two-infall model for the chemical evolution of the Milky Way (Chiappini et al. 1997; Chiappini et al. 2001), which 
relaxes the instantaneous
 recycling approximation but retains the instantaneous mixing approximation, can provide an excellent fit to the majority of
 the data in the Milky Way
 and in particular to the relative abundance ratios in the whole metallicity range (from [Fe/H] =-4.0 to 0).

\item The most important factor in reproducing the [el/Fe] vs. [Fe/H] relations as well as the solar absolute abundances in the 
solar neighborhood is the
 combination of the yields from single low and intermediate mass stars, type Ia and II supernovae. We adopted the star formation
 and infall
 laws that best reproduce the majority of the observations in the Milky Way,
which include other and independent constraints, such as the G-dwarf metallicity distribution, the age-metallicity relation, the current
 amount of gas and fraction of
 stars, the current star formation rate and infall rate as well as the current 
SN rates and their ratio. In other words, we are not allowed to change the star formation rate history, the infall rate history and
 the IMF since they have been tested already on a large amount of observational data.

\item Here, we discuss only elements  for which the contribution of type Ia and II SNe  is relevant. Concerning the yields of SN II we find
 that the yields of WW95 provide the best fit: in particular, no modification is required for the yields of Ca, Fe, Zn, and Ni  as computed
 for a solar chemical composition. For oxygen, the best fit is given by the WW95 yields computed as functions of metallicity.
For the other examined elements (Mg, Si, Ti, K, Sc, Co, Cr and Mn), variations 
of various amounts in the predicted yields are required.   These results are rather
 robust as we ran numerous models spanning a wide range of yields both for SNIa and SNII. However, the numbers shown in Tables 1 and 2 
have to be taken more as median values than definitive values, the aim of this paper beeing to pinpoint in which mass range the yields need
 revision. Another important point  shown in figure 7 to 9 concerns  the relative contributions of SNIa and SNII to the enrichment of the
 Galaxy.
While uncertainties are expected in the predicted yields of the Fe-peak elements  mainly related to the mass-cut (Sc, Cr, Mn, Co), for
 the $\alpha$-elements (Mg, Si, Ti)
it is more difficult to envisage why the yields should be different.
In particular, a common feature, relative to Mg yields in massive stars as
computed by different authors, is the need to substantially increase the 
produced and ejected Mg in stars between 11 and 20$M_{\odot}$ and to decrease it for larger masses. Probably a lower value of the rate
 of the $^{12}C(\alpha, \gamma)^{16}O$ reaction 
 could help in increasing the Mg yields  but it would affect 
the yields of  Si and Ca (Limongi, private communication) or a different treatment of convection.
For K,  the situation is more complicated since there are contributions to this elements also from  neutrino-induced reactions.

\item We tested  the yields for SN II as computed by N97 and LC03 and also in these cases we conclude that modifications are required
 to obtain
 the best fit to the observations.  
 In particular, for the yields of N97, various  modifications 
 are required for all the studied elements;  for the yields of LC03 the O yields should be almost untouched.

\item 
For the yields from type Ia SNe, a revision is needed 
for Mg, Ti, Sc, K, Co, Ni and Zn. In particular, while values  Mg, Ti, Sc,  Zn and Co should 
be larger,
those of K and Ni should be smaller than predicted. Whether these proposed modifications 
are physically plausible  is still to be assessed by the experts in the nucleosynthesis field.

\end{itemize}

\begin{acknowledgements}
We would like to thank Marco Limongi for very useful comments on the 
nucleosynthesis aspects. F.M and C.C. acknowledge financial support from INAF (Italian national institute for astrophysics)
 contract n.2003028039.
P.F. and C.C. acknowledge support  under the ESO visitor program in Garching during the completion of a part of this work.
\end{acknowledgements}

\end{document}